\documentclass[12pt]{iopart}
\usepackage{iopams}
\usepackage{graphicx}
\def\>{\rangle}
\def\<{\langle}
\newcommand{\ket}[1]{|#1\rangle}

\begin{document}

\title[]{Loss Tolerant Linear Optical Quantum Memory By Measurement Based Quantum Computing}


\author{Michael Varnava}

\address{Optics Section, Blackett Laboratory, Imperial
College London, \\London, SW7 2BZ, United Kingdom}

\author{Daniel E. Browne}
\address{Departments of Materials and Physics,
University of Oxford, \\Parks Road, Oxford, OX1 3PH, United Kingdom}

\author{Terry Rudolph}
\address{Optics Section, Blackett Laboratory, Imperial College
London, \\ London, SW7 2BZ, United Kingdom}%
\address{Institute for Mathematical Sciences, Imperial College
London, \\ London, SW7 2BW, United Kingdom}

\begin{abstract}
We give a scheme for loss tolerantly building a linear optical
quantum memory which itself is tolerant to qubit loss. We use the
encoding recently introduced in \cite{varnava} and give a method for
efficiently achieving this. The entire approach resides within the
``one-way" model for quantum computing \cite{brieg}. Our results
suggest that it is possible to build a loss tolerant quantum memory,
such that if the requirement is to keep the data stored over
arbitrarily long times then this is possible with only polynomially
increasing resources and logarithmically increasing individual
photon life-times.
\end{abstract}

\maketitle

\section{Introduction}

Linear optics is a promising candidate for quantum computing.
Photons make excellent qubits. They are very versatile, mobile and
have long decoherence times allowing for data to be confidently
stored in them. Logic gates for linear optical quantum computation
(LOQC) can be built using interferometric linear optical elements
(e.g. phase shifters and polarizing beam splitters), photon
detectors and photon sources in a scalable manner, as shown by
\cite{klm}. Alternative approaches to LOQC using measurement based
computation
 \cite{yoranreznik,NielsenCluster,tezdan} considerably reduce the  overhead
(in extra modes, photons detectors and phase stable circuitry)
necessary for scalable computation. In measurement-based quantum
computation, single qubit measurements alone on entangled
multi-qubit states called cluster-states, or graph-states, implement
the computation. These schemes provided a recipe for efficiently
generating cluster states of arbitrary size using conditional linear
optics and photo-detection. However, in their initial forms they
only succeed if all errors can be ignored.

A cluster state is a multi-qubit entangled state represented
graphically by a graph, where the $n$ vertices of the graph
correspond to qubits prepared in state $\ket{+}$, and the bonds
denote the application of a certain entangling logic gate between
the connected qubits. If we denote by $E(i)$ the set of edges on
this underlying graph connected to vertex $i$, we can compactly
describe such a state in terms of it's ``stabilizer generators'', a
set of operators of the form:
\begin{equation}
X_i \prod_{j\in E(i)} Z_j\textrm{,}
\end{equation}
under which the state is invariant. An operational interpretation of
the stabilizer operators is a prediction of correlations in the
measurement outcomes of certain sets of measurements.

An important property of cluster states is that the application of
parity measurements between qubits not connected on the graph
implements a ``fusion operation'' \cite{tezdan}, whereby the
resultant state is a cluster state which has inherited the graph of
the previous state except that the two nodes representing the
measured qubits have been ``fused'' into a single vertex. The fusion
operation allows one to combine disjoint cluster states and, in
particular, to construct large cluster states from smaller ones. The
fusion operation (and it's linear optical realisation) is the main
tool which will be utilised to build up the entangled states which
will be employed in this paper.

One of the major challenges for implementing LOQC is photon loss.
Photons will only have a finite lifetime, while for quantum
computation quantum information must remain coherent over
arbitrarily long times. Thus a scalable coherent quantum memory is
an important step on the way to developing scalable LOQC. Various
proposals exist for single photon memory involve storing the photon
in optical fibre loops \cite{franson} or in cold atomic clouds
\cite{polzik}.
Our goal is to show that given lossy, single-photon memory devices,
inefficient detectors and inefficient single photon sources, a
memory capable of storing a photonic state indefinitely can be
constructed. The individual photon memory storage times need only
increase logarithmically with the total time required to keep data
qubits in memory.
%
%
Furthermore, this can act as the basis for a gate-based approach to
linear optical quantum computation, which would allow the adoption
of fault-tolerant approaches to correct other non-loss errors
\cite{faultraussen}.

In \cite{varnava} a protocol for loss-tolerant quantum computation
was proposed. At the heart of this scheme was the realisation, that
tree-shaped cluster states can be used as an encoding, each ``tree"
replacing a logical qubit in the un-encoded cluster state. With this
encoding, single qubit losses of up to 50\% can be efficiently
suppressed to yield an effective loss rate for logical qubits which
is arbitrarily close to zero. In this article we describe a
construction procedure for efficiently and loss tolerantly creating
the encoded logical cluster states used for both computation and
memory devices in a linear optical setting and give a full account
of the resources required.

There are two key techniques at the heart the linear optical memory
we propose. The first is the use of specialized cluster states we
term \textquotedblleft hypertrees\textquotedblright. These states
are formed from multiple loss tolerant tree clusters \cite{varnava}
fused together. A nice property of such states is that they allow
(at the level of logical qubits) controlled-phase gates to be
implemented with arbitrary success probability, something which is
not possible via linear optics and measurement on un-encoded
photonic qubits; thus large encoded cluster states can be
constructed, or logic gates can be implemented directly
\cite{divincenzobeenakker}. We expect this technique to be of use
and significance beyond its particular application here.

The second technique is the fact that for the purposes of using
continual teleportation through cluster states to keep a photon
alive, only Pauli measurements are required. This is useful because
it allows for a great amount of parallelization since Pauli
measurements do not need to be adapted based on the outcome of other
measurements. The measurements  can be implemented simultaneously
which helps to relax the requirements of the individual photon
memory, in terms of the amount of time individual photons need to be
stored for. The loss tolerant properties of the tree-structures
employed allow us to attain a higher threshold than other recent
proposals for linear optical based memory \cite{ralphmem}.

We point out that here we only address detected losses (erasures),
since these form the dominant errors we should expect within LOQC.
Other work has addressed LOQC within the context of undetected
errors, see e.g.\cite{terhal,dawsonhasel}. Furthermore, the
near-deterministic logic gates this scheme allows on the level of
encoded qubits could allow the implementation of error-correction
schemes for a wider variety of errors \cite{faultraussen}.

The paper is structured as follows: First we give a brief outline of
the loss-tolerant approaches in \cite{varnava}. We then give a
resource efficient strategy for creating the trees used in the
encoding. After this, we  introduce a scheme for joining
tree-encoded qubits in an asymptotically deterministic way by employing
``hypertree'' structures. Later on we will give an account of how
one can build the loss tolerant quantum memory with the properties
claimed earlier. A full resource count will be provided throughout
to demonstrate that the scheme introduced is resource efficient.

\section{A resource for loss-tolerant computation}

\begin{figure}[ht]
\centering
\includegraphics[width=8cm]{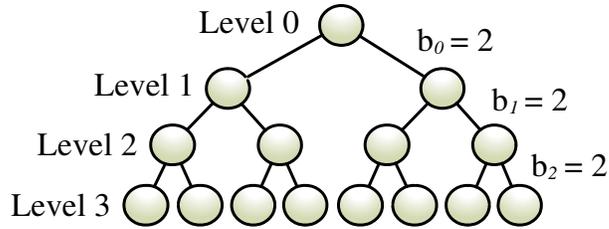}
\caption{\label{regtree} A tree cluster state with branching
parameters: $\{b_{0},b_{1}...b_m\}=\{2,2,2\}$ .}
\end{figure}

In \cite{varnava} a protocol is outlined in which cluster states
with a tree-structure are used to encode qubits to enable
loss-tolerant measurement-based quantum computation. An example of a
tree-cluster state a state is shown in Figure \ref{regtree}. Tree
cluster states are fully specified by their branching parameters,
$\{b_{0},b_{1}... b_m\}$; for example $b_{i}$  equal to the number
of branches coming down from each qubit in level $i$. When each
qubit of a logical cluster state is encoded by a tree cluster state,
then a plethora of alternative measurement patterns  become
available for implementing the desired logical operation; namely the
measurement of the original single qubit in some arbitrary basis.

The key idea is that one can actively change this measurement
pattern as one goes along to adapt for lost qubits detected on the way.
 At instances where qubit measurements fail, then the special quantum correlations present on
the tree cluster states can be exploited to allow the outcome of
measurements on the lost qubits to be inferred by measuring other
qubits on the tree, which due to the entanglement in the state will
be correlated with the lost outcome. The logical operation can thus
proceed with an alternative measurement pattern which is still
available.

In \cite{varnava} we showed that provided the trees have sufficient
branching, independent qubit loss errors at rate $\epsilon$ can be
tolerated for any $\epsilon<0.5$. More precisely, with only a
polylogarithmic scaling of the number of qubits, $Q$, required to be
present on a tree, the effective loss rate, $\varepsilon
_{\small{\textrm{eff}}}$, is exponentially rapidly reduced to zero.
More recently \cite{varnava2} we showed that this threshold for
$\epsilon $ can be translated into an LOQC\ architecture with the
requirement that the product of the detector efficiency, $\eta
_{D}$, and the single photon source efficiency, $\eta _{S}$, has to
be greater than 2/3.

\section{Creating the Tree Clusters efficiently.}

The special tree cluster states introduced in \cite{varnava} are
fully specified by the branching parameters $b_{0}$ to $b_{m}$\ as
they are traversed from the top to the bottom levels.
We review briefly in this section an efficient strategy for building
these trees using \ redundantly encoded ``2-trees" as the primitive
building block and fusing them together into larger cluster states
using the Type-II fusion gate. Type-II fusion is a variant of the
fusion operation which can be employed when (at least) one of the
qubits acted upon is ``redundantly encoded''. Redundant encoding is
the simplest form of coding one could imagine. The logical state
$\ket{0}$ is represented by $n$-qubits in state $\ket{0}$, i.e.
$\ket{0}^{\otimes n}$ and $\ket{1}$ is represented by
$\ket{1}^{\otimes n}$. It is straight-forward to confirm that a
Bell-state projection between such a pair of qubits acts as a parity
measurement \cite{pittman} - realising a fusion operation. A Type-II
gate is a linear optical realisation of such a Bell-measurement. It
is effected by the combination of a polarizing beamsplitter oriented
at $45^\circ$, followed by number-resolving and
polarization-resolving detectors on both output modes. Here we will
use a slightly modified version of the gate by inserting a
$45^\circ$ polarization rotator on each of the two spatial modes
prior to the beamsplitter. For the case where two photons are
detected at the same detector, the gate fails and the effect is to
measure the input qubits in the Z basis (instead of in the X basis
as in the original version of the gate proposed in \cite{tezdan}).
The gate also fails when, less than 2 photons are detected in total
by the gate (because of loss, detector inefficiencies etc.). The
gate is only deemed ``successful'' (i.e. the desired fusion
operation is implemented) when one and only one photon is detected
in each output spatial mode.

In the ideal case, where we assume no qubit loss is present and
perfect sources and detectors are available, the success probability
rate for the linear optical Type-II gate acting on photons (which
are in a locally maximally mixed state - as is the case for cluster
state photons) is 50\%. In a more realistic scenario, however, the
actual success rate, $P_{II}$, for the Type-II gate is compromised
by the detection efficiencies $\eta_D$ of the two detectors and the
independent loss probability $\epsilon$ of the two photons present
in the gate. Since \emph{both} photons \emph{must} be present and
\emph{both} detectors \emph{must} detect a photon then $P_{II}$ is
reduced to $\frac{\left( 1-\epsilon \right) ^{2}\eta _{D}^{2}}{2}$.

\begin{figure}[ht]
\centering
\includegraphics[width=12cm]{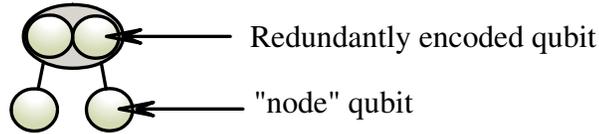}
\caption{\label{2-tree} A 2-tree is a 3 qubit cluster state with a
centrally redundantly encoded logical qubit which branches out to 2
``node" qubits.}
\end{figure}

\begin{figure}[ht]
\centering
\includegraphics[width=12cm]{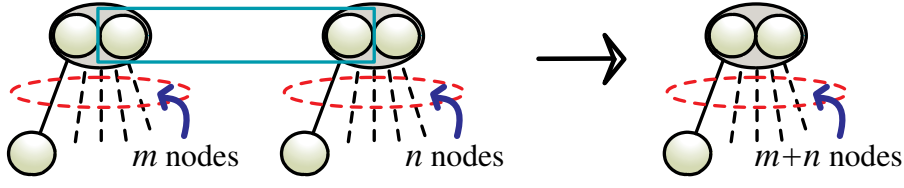}
\caption{\label{ntreejoin} Successfully Type-II fusing an n-tree
with an m-tree creates an n+m-tree. The 2 photons used by the
Type-II gate are indicated by the green box. }
\end{figure}

Generally we define an \textquotedblleft n-tree\textquotedblright\
as consisting of a central redundantly encoded qubit (in 2 physical
qubits with logical bases $\left\vert {00}\right\rangle$ and
$\left\vert {11}\right\rangle$), to which $n$ node qubits are connected on the graph. The example of Figure \ref{2-tree} shows a
2-tree.
The strategy we follow is to build the trees from bottom to top
adding levels of qubits in the following way: First we fuse
$2$-trees together to form $b_{m}$-trees. This is achieved through a
series of post-selection steps. First we post-select upon successful
fusion attempts to create a resource of $4$-trees from joining
$2$-trees together. Then we fuse 4-trees together and create a
resource of $8$-trees subject to successful type-II fusions and so
on. Generally we fuse $m$-trees with $n$-trees and upon successful
outcomes on the Type-II detectors we obtain $(m+n)$-trees (see
Figure \ref{ntreejoin}).
The expected number of $2^{l-1}$-trees required to create a
$2^{l}$-tree is equal to $2/P_{II}$. Thus the expected number of
$2$-trees required to build a single $2^{l}$-tree is $\left[
2/P_{II}\right] ^{l-1}$.
Furthermore, it can readily be seen that in order to create a $b_{m}$-tree such that $%
2^{l-1}\leq b_{m}\leq 2^{l}$, then on average the number of
$2$-trees required is $\leq \left[ 2/P_{II}\right] ^{\log _{2}\left( b_{m}\right) }=$%
poly$\left( b_{m}\right)$. In this way we can efficiently create the
lowest level of the desired trees with the branching parameter
needed to tolerate the given loss rate.

\begin{figure}[ht]
\centering
\includegraphics[width=12cm]{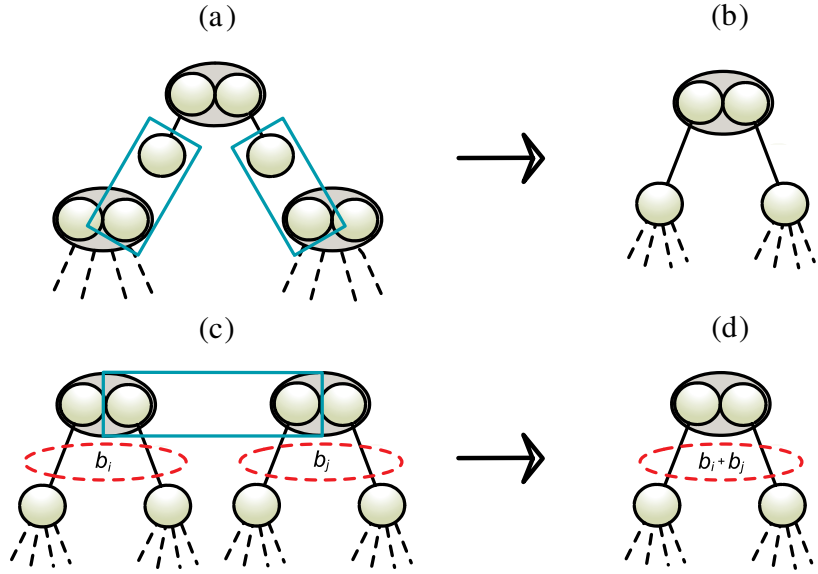}
\caption{\label{addlevel} Adding a new level requires 2 steps. 1:
(a) $\rightarrow$ (b) First fuse a 2-tree with 2 of the existing
trees to add a new higher level. 2: (c)$\rightarrow$ (d) Fuse states
created in (b) together to increase the branching at the added
higher level.}
\end{figure}

There are two steps involved for each additional level we would like
to add. First we use 2 successful $b_{m}$-trees created earlier and
fuse them together with a 2-tree in the fashion shown in Figure
\ref{addlevel}(a) which uses two Type-II gates. Upon successfully
performing the gates the resulting cluster state is the one shown on
Figure \ref{addlevel}(b).
This is now a tree with branching parameters
$\{b_{0},b_{1}\}=\{2,b_{m}\}$. The second step is to fuse these
trees together as shown in Figure \ref{addlevel}(c) to increase the
top level branching from 2 to $b_{m-1}$. We can now increase the
branching parameter on the top level from 2 to b$_{m-1}$ by
combining these tree clusters together, much as we combined the
initial 2-trees. To complete the first step, the expected number of
2-trees required in order to create a single tree with branching
parameters $\{2,b_{m}\}$ is $(2$poly$\left( b_{m}\right)+1) \left[
1/P_{II}\right] ^{2}$. To complete the second step, the expected
number of trees with branching parameters $\{2,b_{m}\}$ required in
order to create a single tree with branching parameters
$\{b_{m-1},b_{m}\}$ is $\leq \left[ 2/P_{II}\right] ^{\log
_{2}\left( b_{m-1}\right) }$. Therefore the overall expected cost in
2-trees required to create one such
tree is $\leq \left[ 1/P_{II}\right] ^{2}$poly$\left( b_{m-1}\right) $poly$%
\left( b_{m}\right) $. This suggests that the extra added level with
branching
parameter $b_{m-1}$ incurs an increasing factor $\left[ 1/P_{II}\right] ^{2}$%
poly$\left( b_{m-1}\right) $ in the 2-trees overhead. Iterating the
process in order to add all required levels suggests that in order
to create one
tree cluster state with the full branching parameter profile $%
\{b_{0},b_{1}...b_{m}\}$ (as required in \cite{varnava}) then the
expected number of
2-trees required satisfies:
\begin{equation}
  \left\langle N_{2-trees}\right\rangle \leq \left[
\frac{1}{P_{II}}\right]
^{2m}\prod\limits_{i=0}^{m}\textrm{poly}\left( b_{i}\right)
\textrm{.}
 \end{equation}
The overall conclusion is that the expected number of qubits
consumed in order to build a tree containing $Q$ qubits is
polynomial in $Q$, since $m\leq \log _{2}\left( Q\right) $.

\section{From trees to ``hypertrees''.}

\begin{figure}[ht]
\centering
\includegraphics[width=12cm]{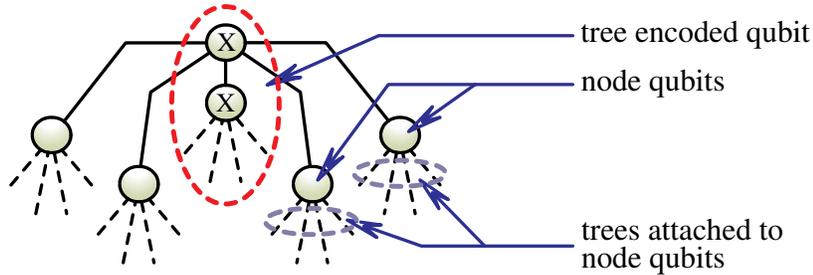}
\caption{\label{hypertree} A hypertree consists of node,
tree-encoded qubits (used in Type-II fusions for joining hypertrees)
which are attached on a central (circled) tree encoded qubit
intended for a logical cluster state.}
\end{figure}

In this section, we shall introduce a new cluster state structure
which we call a ``hypertree''. In comparison to the tree-clusters
introduced in \cite{varnava}, these have useful extra properties
which we shall describe below. An example of a hypertree can be seen
in Figure \ref{hypertree}. Hypertrees are similar to the original
trees, the only differences being the addition of an extra higher
level. We assume that two of the qubits have been successfully
measured in the X basis. The hypertree state is the state after
these measurements have been performed. We retain them to simplify
the states description. In practice, one would generate the
post-X-measurement hypertree state directly.

Each hypertree must be thought of as being a single, tree-encoded,
logical qubit which is directly linked to a number of node qubits.
Each of these node qubits are the root of a further tree structure.
These node-qubits will be used as the input of Type-II fusion gates
to join together logical tree-encoded qubits (directly linked with
them within their hypertrees) into larger computation-specific,
tree-encoded cluster states. These node qubits serve the same role
as the leaf node qubits introduced by \cite{dawsonhasel} however the
trees attached to these node qubits allow them to be measured
indirectly and loss-tolerantly allowing one to recover from failures
of the fusion gate. An alternative description of the hypertree
structures (as redundantly encoded qubits which are further tree
encoded) was presented in \cite{varnava2}.

As we shall see later, the node qubits provide a number of different
alternatives whereby one can attempt to join two logical qubits
together. At most \emph{one and only one} Type-II gate is required
to succeed between the node qubits of any two distinct hypertrees in
order for the logical tree-encoded qubits to be successfully joined
together. This entire process is analogous to a logical
Controlled-Phase (CZ) gate performed between the logical qubits.
This is an essential step in creating the computation-specific,
tree-encoded cluster state to be used by a computation. Further on
we will see that the reason for going through the intermediate steps
of first building hypertrees and then Type-II fusing their node
qubits together in order to build computation-specific cluster
states is that it allows us to join logical tree-encoded qubits
together in a near-deterministic fashion by using the probabilistic
Type-II fusion gates; and that this is possible with just polynomial
resource overheads.
\begin{figure}[ht]
\centering
\includegraphics[width=12cm]{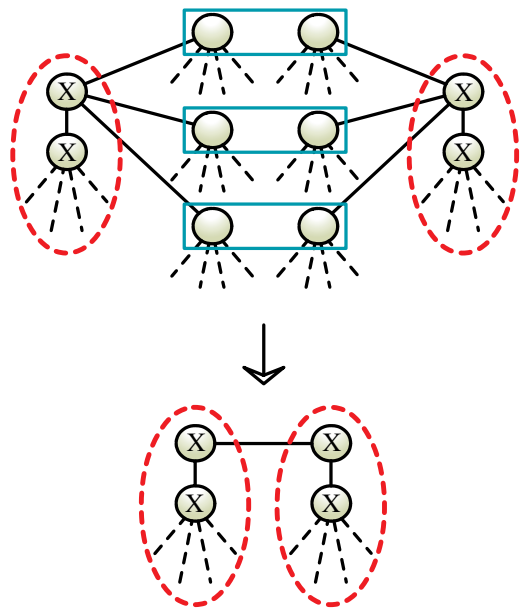}
\caption{\label{hyperjoin} Two hypertrees can be joined together by
Type-II fusing together their node qubits. Only one Type-II is
required to succeed. The overall effect is to create a CZ bond
between the 2 tree-encoded qubits present in the original hypertrees
if indeed at least one of the Type-II gates succeeds.}
\end{figure}

In Figure \ref{hyperjoin} we show how two hypertrees can be linked
together using Type-II gates. To see why it is we only require one
Type-II gate to succeed we need to closely examine all the possible
outcomes a Type-II gate can give and explain how they can be dealt
with. A Type-II gate has 3 distinct sets of outcomes: Either (a)
only one or no photons will be detected (because of loss or detector
inefficiency) or (b) both photons will be detected at the same
detector or (c) both photons will be detected, one at each separate
detector. From these possibilities only (c) is accepted as the
correct outcome. The outcomes (a) and (b) would be catastrophic if
encoded qubits are not used. However, the fact that here there is a
tree joined on every node photon means that we can execute specific
measurement patterns on those trees to rectify any of the possible
outcomes with arbitrary success probability. In particular, if
outcome (a) occurs and the measured qubits are lost, then they can
be indirectly and loss tolerantly measured in the Z-basis by
measuring qubits in their attached tree as was discussed in
considerable detail in \cite{varnava}.

If outcome (b) occurs then this has the effect of measuring the node
qubits in the Z basis. This is the least damaging result for an
unsuccessful outcome, as it simply removes the node qubits from the
two hypertrees. This is precisely the reason for using the modified
version of the Type-II gate mentioned earlier, as in cluster state
computation the effect of Z measurements is to remove the measured
qubits from  the cluster state. Note that measuring the remainder of
the connected tree can be advantageous since the extra measurements
can provide additional information as to what the Z measurement
outcome on the node qubits should be. Obtaining many such
\textquotedblleft votes\textquotedblright\ for a given outcome and
applying a majority voting over these results can greatly suppress
logical errors such as depolarisation \cite{ralph,varnava3} although
a full discussion of this effect is beyond the scope of this
article.
\begin{figure}[ht]
\centering
\includegraphics[width=14.5cm]{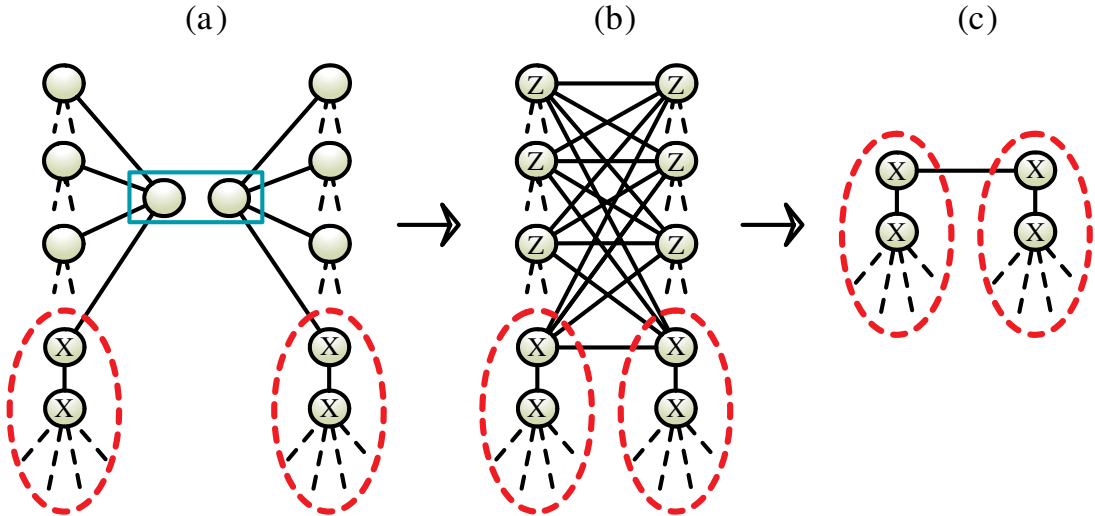}
\caption{\label{TIImess} (a) A Type-II gate is implemented between
two node photons of two distinct hypertrees. (b) The resulting state
after the successful Type-II outcome. (c) Resulting state after
measuring the undesired qubits in the Z basis. This is now a state
whereby the two logical qubits are successfully linked by a CZ bond.
}
\end{figure}

Finally, if outcome (c) occurs then we know that the gate has been
successfully implemented. Figure \ref{TIImess} shows explicitly an
example of a successful Type-II gate. Once the successful outcome is
received then there are a number of new bonds created between the
two hypertrees as it is shown on Figure \ref{TIImess}(b). Of all
these new bonds, only the direct bond between the two logical qubits
is required. Any of the other bonds emerging from the qubits, that
used to be in the first level of the trees attached onto the
original node qubits from either hypertree, must now be removed.
This can be achieved by measuring all these qubits in the Z basis.
Note that these Z measurements can again be implemented with a
success probability arbitrarily close to unity, because they can
also be effected indirectly. Remember that these measurements are
effected on qubits each of which was at the top level of a tree.
Thus the Z measurements can also be effected indirectly by following
measurement patterns on the lower levels of these trees in the
fashion explained by \cite{varnava}.

It is clear, therefore, that regardless of which Type-II fusion
outcome occurs there is a specific measurement pattern that can be
followed to deal with it. The purpose of these hypertrees is to
(asymptotically) deterministically join tree-encoded qubits together
using lossy and probabilistic Type-II gates. At least one Type-II
gate must succeed in order to be able to join two logical qubits
together. As such it is expected that the higher the number of node
qubits present in hypertrees, the higher the effective probability
for at least one Type-II fusion gate to succeed. We now analyse in a
little more detail the requirements for resource efficiency.

The computation-specific cluster states, used in the one way model
for quantum computing \cite{brieg}, can be thought of as being
created in two steps. First the qubits are initiated in the
$\ket{+}$ state and then the bonds present in the cluster state are
formed by effecting controlled-phase gates between pairs of qubits.
Suppose we would like to build a computation-specific cluster state
formed by tree-encoded logical qubits. Such a cluster state can be
built with arbitrary success probability by first initiating
hypertrees and then fusing those together. Recall that hypertrees
consist of tree-encoded logical qubits attached to node photons
(which in turn have a tree attached on them). We showed above that
Type-II fusing node qubits of two distinct hypertrees has the effect
of forming a direct CZ bond between the tree-encoded logical qubits
present in these hypertrees. More importantly is that the
probability with which this bond is effected can be increased
dramatically, simply by allowing for a large number of node qubits
to be available on each of the hypertrees containing the logical
qubits. This is because that would allow for the possibility of a
large number of Type-II attempts to be implemented between the node
qubits of the two hypertrees. Since the requirement is just
\emph{one} of those fusion attempts needs to succeed, the effective
success probability for joining the logical hypertrees together is
increased.

Assume w.l.o.g that any logical qubit in the above computational
cluster state must be bonded to $n$ other logical qubits. Further
assume that for any such bond we would like to allow for a maximum
of $k$ Type-II fusion attempts to be performed. This suggests that
we would want to use hypertrees which have $kn$ node photons. To
build such hypertrees would require an expected number of $\left[
1/P_{II}\right] ^{2}$poly$\left( kn\right) $. To see this remember
that hypertrees are in effect identical to the regular trees with an
additional higher level with branching factor $kn$.

On the other hand, the probability for successfully joining 2
tree-encoded logical qubits together (using their hypertrees) is
given by:
\begin{equation}\label{PZ}
P_{CZ} =\left[ 1-\left( 1-P_{II}\right) ^{k}\right] P_{tree}^{2k}
\textrm{.}
\end{equation}%
Here $P_{tree}$ is the probability for successfully implementing the
necessary measurement pattern on the tree attached to a node photon
as soon as the result of the Type-II fusion gate involving the node
photon becomes available. There are $2k$ such node photons involved
with every attempt to fuse 2 hypertrees together and the
$P_{tree}^{2k}$ factor is present in the expression for $P_{CZ}$
above because all measurement patterns that have to be followed on
the trees attached on these $2k$ node photons must succeed in order
for the successful fusion of the hypertrees.

The objective is to check whether $ P_{CZ}$ can approach unity with
efficient resource scaling. Consider first the factor $\left[
1-\left( 1-P_{II}\right) ^{k}\right]$ in the expression for $
P_{CZ}$. The success probability for performing a Type-II gate,
$P_{II}$ is a fixed, physical parameter of the experimental setup;
thus one can choose a value for $k$ to compensate for any value of
$P_{II}$ efficiently. Here what we mean by efficiently is that even
with a very modest linear increase of the value of $k$ the factor
$\left[ 1-\left( 1-P_{II}\right) ^{k}\right]$ increases and
approaches unity exponentially fast no matter how small $P_{II}$ is.

However, this linear increase in the value of $k$ will have a
noticeable effect on the second factor in the expression for
$P_{CZ}$ given by $P_{tree}^{2k}$. In \cite{varnava} we showed with
numerical analysis that $P_{tree}$ is related to $Q$, the number of
physical qubits present in a tree encoded logical qubit, by the
expression:
\begin{equation}
\log \left( Q\right) =c\log \log \left( \frac{1}{1-P_{tree}} \right)
\textrm{, where } c\approx 4.5.
\end{equation}
Rearranging gives: $ P_{tree}= 1-\exp \left( -Q^{1/c}\right)$ thus:

\begin{equation}
P_{tree}^{2k}\simeq 1-2k\exp \left( -Q^{1/c}\right)\textrm{,}
\end{equation}
is a good approximation since $1\gg \exp\left(-Q^{1/c}\right)$ even
for very modest values of $Q$.

From this we can deduce that $P_{tree}^{2k}$ is linearly decreasing
with $k$, but the effect can be over-compensated by the choice of
$Q$ since $P_{tree}^{2k}$ is \emph{exponentially} dependent on
$Q^{1/c}$. By \emph{linearly} increasing $Q^{1/c}$, one can
over-compensate the effect of the previously chosen value for $k$
and still have $P_{tree}^{2k}$ approaching unity
\emph{exponentially} fast.

We conclude therefore, that $P_{CZ}$ can approach unity
exponentially fast with just \emph{linearly} increasing $k$ and
\emph{polynomially} increasing $Q$ with respect to $P_{CZ}$. This is
an efficient resource scaling as the number of qubits present on a
hypertree with say $nk$ node qubits, contains $nk(Q+1)$ physical
qubits in total. Hence the overall resource scaling is polynomial
with highest degree equal to $c+1$ with respect to $P_{CZ}$.

%
%
%
\section{A loss Tolerant Quantum memory}

\begin{figure}[ht]
\centering
\includegraphics[width=12cm]{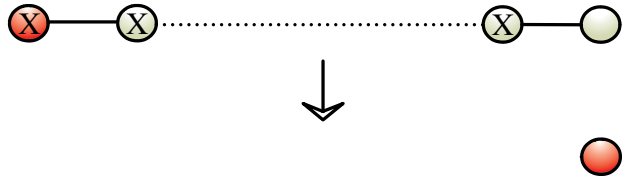}
\caption{\label{Teleport} Teleporting a data qubit through a longer
tree-encoded linear cluster.}
\end{figure}

Using the hypertrees introduced above one can create linear clusters
of tree-encoded qubits. Such linear clusters and measurements in the
X basis can then be used as a loss tolerant quantum memory for the
one way model for quantum computing.
\begin{figure}[ht]
\centering
\includegraphics[width=12cm]{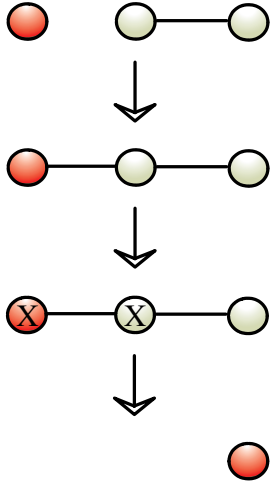}
\caption{\label{memory unit} The quantum memory proposed works in a
teleportation approach. First the tree-encoded data qubit (red)
which is in an arbitrary state $\alpha \left\vert 0\right\rangle
+\beta \left\vert 1\right\rangle $ is joined on a tree-encoded
linear cluster. By performing logical X measurements on the data
qubit and the next on its right teleports the state on the qubit
furthest to the right.}
\end{figure}
The memory we propose works in a teleportation-type approach. As can
be seen on Figure \ref{memory unit}, the main idea is to join a data
qubit with a linear cluster of 2 qubits. Subject to successfully
achieving this, one can proceed by measuring the original data qubit
and the first qubit of the former 2-qubit linear cluster in the X
basis. Subject to successfully implementing these steps, the state
of the original data qubit has now been teleported to the last
qubit, (formerly the second qubit of the 2-qubit linear cluster).
One can of course iterate this process for as long as necessary to
store the data qubit. This in fact is exactly analogous to joining a
longer linear cluster in the first place and performing an even
number of X measurements (see Figure \ref{Teleport}); the effect is
to teleport the data qubit through a longer cluster, but equally it
can be argued that the the effect is to store the data qubit for a
longer period of time.
One can deduce that the method proposed here is resource efficient
way for constructing a quantum memory. In the previous section we
showed that with just polynomially increasing resources, one can
perform logical CZ-gates between tree encoded qubits with
exponentially increasing success probability, $P{CZ}$. In addition,
the results of \cite{varnava} indicate that the effective success
probability, $P_{tree}$,  for performing a measurement on a tree
encoded qubit, can be exponentially increased towards unity by
polynomially increasing $Q$. These two are the operations required
for the proposed memory.

Suppose we wish to create a memory that stores qubits for a time
$\tau_{mem}$ with an overall success probability $P_{mem}$. The
method we will actively create and operate the memory would be as
follows:\\ \\    \par~~~1. Create a new hypertree.\par~~~2. Perform
a logical CZ-gate between the data qubit and the new
hypertree.\par~~~3. Measure original data qubit in X basis.\par~~~4.
Label the remaining logical qubit as the new data qubit and repeat
from 1. \newline
\\
Suppose also that the time it takes for one cycle (steps 1 to 4 to
complete) is $\tau_q$. Note that the overall success probability for
performing one cycle is given by $P_{CZ}P_{tree}$. In other words it
is the probability of successfully joining the newly created
hypertree to the data qubit followed by successfully measuring the
original data qubit in the X-basis. This would suggest that:
\begin{equation}
P_{mem}=(P_{CZ}P_{tree})^\frac{\tau_{mem}}{\tau_q}\textrm{,}
\end{equation}
as we would need to repeat the cycle $\frac{\tau_{mem}}{\tau_q}$
times in order to store a data qubit for a period of $\tau_{mem}$.
(Incidentally the number of cycles has to be even in order to
perform the identity gate which is what in effect the memory gate
actually is in this setting, however this feature does not affect
the resource scaling calculations that follow.)

By substituting Eqn.~(\ref{PZ}) for $P_{CZ}$ the expression for the
memory success probability becomes:
\begin{equation}
P_{mem}\approx \left[ 1-\left(\frac{\tau_{mem}}{\tau_q}\right)\left(
1-P_{II}\right) ^{k}\right]\left[  1-
\left(2k+1\right)\left(\frac{\tau_{mem}}{\tau_q}\right)\exp
\left(-Q^{1/c}\right) \right]
\end{equation}
With a bit of thought one can see that $k$ and $Q^{1/c}$ scale
logarithmically with $\tau_{mem}$. To see this suppose we need to
find $k'$ such that:
\begin{equation}
\left(\frac{\tau_{mem}}{\tau_q}\right)\left( 1-P_{II}\right)
^{k'}=\left( 1-P_{II}\right) ^{k} \textrm{.}
\end{equation}
Taking logarithms on both sides gives:
\begin{equation}
k'=k -
\frac{\log{\left[\frac{\tau_{mem}}{\tau_q}\right]}}{\log{\left[1-P_{II}\right]}}\textrm{.}
\end{equation}
Similarly, suppose we wish to find $Q'$ such that:
\begin{equation}
\left(\frac{\tau_{mem}}{\tau_q}\right)\exp
\left(-Q'^{1/c}\right)=\exp \left(-Q^{1/c}\right)\textrm{.}
\end{equation}
Taking logarithms on both sides gives:
\begin{equation}
Q'^{1/c}=Q^{1/c}+\log{\left[\frac{\tau_{mem}}{\tau_q}\right]\textrm{.}}
\end{equation}
Clearly by logarithmically increasing both $k$ and $Q^{1/c}$ with
respect to the memory time, $\tau_{mem}$, has the effect of
increasing the memory success probability to:
\begin{equation}
P_{mem}=(P_{CZ}P_{tree})^\frac{\tau_{mem}}{\tau_q} \rightarrow
P_{CZ}P_{tree} \textrm{.}
\end{equation}

Such a memory will require $\tau_{mem} /\tau_{q}$ hypertrees in
order to store a data qubit for a time $\tau_{mem}$. Thus overall,
resources scale proportionally to
$\left(\frac{\tau_{mem}}{\tau_{q}}\right)\left[\log\left(\frac{\tau_{mem}}{\tau_{q}}\right)
\right]^2$.  The resource scaling here, is with regards to the total
time $\tau_{mem}$ with which the qubit is required to be stored.
With regards to the success probability rate, $P_{mem}$, by which
the data stored is stored over $\tau_{mem}$ , the results of the
previous section for the resource scaling with respect to $P_{CZ}$
imply that $P_{mem}$ can increase exponentially fast towards unity
with similar polynomially increasing resources. $P_{mem}$ differs
from $P_{CZ}$ by a mere factor of $P_{tree}$ (after considering the
resource scaling with respect to the $\tau_{mem}$) suggesting that
the resource scaling with respect to $P_{mem}$ would be polynomial
with degree $c+1$ which is very similar to the resource scaling with
respect to $P_{CZ}$ discussed in the previous section.

As we now explain, the fidelity of the quantum memory we are
proposing can be defined as the success probability of the memory.
This of course is only true under the assumptions we made throughout
this article namely that the only source of error is loss due to
imperfect detectors, imperfect single photon sources and lossy
components. We also assume that no dark counts occur at the
detectors and that the single photon sources do never emit 2-photon
states. Under this model, the Type-II gates filter out all possible
outcomes by discarding any input states that gave rise to an
erroneous outcome as soon as such outcomes become known. Conversely
this suggests that whenever a hypertree is postselected subject to
successful outcomes \emph{on all} the Type-II gates involved in its
preparation then such a state may be regarded as being prepared
perfectly.

The (yet) unmeasured qubits of the hypertree may not all, have been
present during the preparation of the state and thus may not have
acquired the relevant entangling bonds intended by the Type-II
gates. Such lost qubits would inevitably fail to be detected when
their measurement is attempted and the protocol proposed in
\cite{varnava} can deal with such instances. However the important
point to note is that the Type-II gates have the property of taking
imperfect source states at the input (i.e states with lost photons
prior to the input of the Type-II gate, but no loss from the pair of
photons operating the gate) and producing output states (supposing
the correct Type-II gate outcome) which are identical to states that
are produced by perfect input states which undergone loss of the
\emph{same} qubits only \emph{after} the action of the Type-II gate.
In other words if we were to model loss by a beamsplitter of
reflectivity $\eta$ placed at each input spatial mode of a Type-II
gate, we find that we can commute the two beamsplitters to the two
output spatial modes of the gate prior to the detectors. This is
specifically true whenever the Type-II gate is operated by \emph{at
most one} photon in each of the input modes which is indeed always
the case in the construction of the memory. The property of the
Type-II gate just described implies that the fidelity of the states
created using this approach are only affected by loss. Thus the
probability by which a memory can succeed also gives the fidelity of
the physical quantum state constituting the memory.

\section{For how long do the memory photons need to be stored?}

We will give an estimate on the maximum time, $\tau _{\max }$,
individual photons in the memory resource need to be stored for in
terms of the time, $ \tau _{II}$, it takes for a Type-II gate and
associated classical feed-forward to complete (essentially the
number of steps in the protocol). In order to simplify the
derivation we are also assuming that $ \tau _{II}$ is the time
required to perform single qubit measurements and the associated
classical feed-forward, although it must be appreciated that in
reality such measurements could take slightly more time that the
Type-II gates. However the vast majority of the time steps involved
in the building process of the quantum memory only involve Type-II
fusion gates for the creation and joining of the hypertrees. Thus if
the time required to perform single qubit measurements is comparable
with $\tau _{II}$, it should not make a significant difference in
the estimate derived for $\tau _{max}$. In giving this estimate, we
make the assumption throughout that the resources for implementing
parallel computations are available in every step. $\tau _{\max }$
is thus the time it takes from the moment individual un-entangled
photons are produced until they are finally measured as part of the
linear clusters used in the memory.

We estimate this time to be
\begin{equation}
\tau _{\max }=\left[ \sum\limits_{i=0}^{m}\log _{2}\left(
b_{i}\right) +m+\log _{2}\left( kn\right) +C\right] \tau _{II},
\end{equation}
where $b_{i}$ are the branching parameters and $m$ is the maximum
depth of the trees cluster states introduced in \cite{varnava} and
$C$ is a constant $\sim 5-8$.

To derive this expression for $\tau _{\max }$ we count first the
time steps required to build 2-trees out of un-entangled photons,
then the number of time steps it takes to build trees out of
2-trees, then the time it takes to build trees into hypertrees and
lastly the time it takes to implement all the Type-II fusion gates
along with the single photon measurements, to join together tree
encoded qubits as linear logical clusters and measure the logical
qubits.

The time it takes to build 2-trees from un-entangled photons is
equal to $ 2\tau _{II}$. One $\tau _{II}$ time step is required to
build the intermediate three photon GHZ states, and another $\tau
_{II}$ is required to fuse those into 2-trees.

To see what the total time is to build the trees introduced in
\cite{varnava} using 2-trees we need to note first the number of
$\tau _{II}$ time steps required in order to increase the branching
at any level from 2 to $b_{i}$ (see Figure \ref{addlevel} step 2).
At each $\tau _{II}$ time step we attempt fusion gates in order to
join trees together to double the top level branching by post
selecting the successful Type-II fusion gate outcomes. Thus it takes
approximately $\log _{2}\left( b_{i}\right) \tau _{II}$ time steps
to increase the branching to $b_{i}$. To add a higher level on the
existing sub-trees with branching equal to 2 (see Figure
\ref{addlevel} step 1) requires one $\tau _{II}$ time step. Thus
overall the number of $\tau _{II}$ time steps required to build
trees from 2-trees is $ \sum\limits_{i=0}^{m}\log _{2}\left(
b_{i}\right) +m $.

To build trees into hypertrees essentially means that we want to add
an additional higher level with branching equal to $kn$. Thus by
following the same logic this can be achieved by $\log _{2}\left(
kn\right) $ extra $\tau _{II}$ time steps.

In order to implement fusion gates on hypertrees in order to join
their tree-encoded logical qubits into tree-encoded linear clusters
(as required by the proposed memory gate), requires merely 3 $\tau
_{II}$ time steps. This is because \emph{all} the Type-II gates can
be implemented simultaneously in one $\tau _{II}$ time step.
Provided at that least one of these gates is successful (which
occurs with near unit probability) the desired fusion between the
encoded logical qubits can be engineered by choosing appropriate
measurement patterns for the subtrees attached to these node qubits.
Whatever the outcomes of each of these gates will be, the
measurement pattern that dictates what would has to be performed on
the trees attached to each of the node photons on all the hypertrees
involved, would be known as soon as the fusion outcome is
registered. These measurement patterns would take at most 2 $\tau
_{II}$ time steps to complete. This is because normally we can
attempt to measure in one $\tau _{II}$ time step all the qubits in
level 0 of the trees \cite{varnava} attached on every node photon.
Then, subject to whether or not the measurements on this level
succeed or fail because of loss, this would define a distinct
measurement pattern that must be implemented on all the remaining
qubits of the lower levels of the tree. This measurement pattern
gives the basis in which each of the remaining qubits in the trees
attached to the node qubits has to be measured. The measurement
bases of these patterns are all Pauli measurements and are not
dependent upon the patterns of loss within them. Therefore this
entire set can be measured in one time step.

The last thing remaining is to perform the logical X measurements on
the data qubit, and the adjacently joined qubit from the linear
cluster (see Figure \ref{memory unit}), remembering that both these
are tree encoded. In order to implement the logical X measurements
would require a set of many physical measurements \cite{varnava}.
However all these measurements can be performed in two $ \tau _{II}$
time steps. First we attempt X measurements on all of the physical
qubits at level 0 of the trees in both of these logical qubits. As
before, depending on whether or not loss occurs in the measurements
defines a distinct measurement pattern that can be implemented on
all the remaining qubits of the tree. This again can be implemented
in one further $\tau _{II}$ time step because all the measurements
are again of Pauli observables.

Note that the expression for $\tau _{\max }$ is logarithmically
dependent on the branching parameters of the trees and hypertrees
used for the encoding and creation of the logical cluster states.
This suggests that if there are enough resources available to allow
for any operations to be performed in parallel  this loss tolerant
quantum memory is very fast,  relying on qubits which do not have to
be stored over long times.

\section{Individual photon memory}

In the previous sections we assumed that photons not used by a
Type-II fusion gate during the creation of the quantum memory can be
perfectly stored until the memory is created. Of course, this
assumption is not reasonable in a laboratory implementation.

Suppose that $P_{\tau_{II}}$ is the probability of successfully
storing a photon not used in a Type-II for a time $\tau_{II}$.
Further assume the pessimistic scenario where every photon (used in
the building process of the quantum memory we are proposing) had to
survive for the maximum time $\tau_{max}$. This would suggest that
the probability of successfully storing any photon would be
$\left(P_{\tau_{II}} \right)^{ \tau_{max}/\tau_{II}}$.
%
%
%
%
%
%
Here we make the assumption that the individual photon memory is
similar in form to the cyclical quantum memory for photons proposed
in \cite{franson} (i.e. the rate of photon loss during storage is
constant). In other words the probability of storing the photon
degrades by a factor of $P_{\tau_{II}}$ for every $\tau_{II}$
time-step the photon is stored.

In \cite{varnava} it was shown that it is possible to perform
universal quantum computing using tree encoded qubits, provided that
the probability of successfully detecting the physical qubits on the
trees is greater than 50\%. This implies that:
\begin{equation}\label{ineq}
(1-\epsilon)\eta_D \left(P_{\tau_{II}}\right) ^{
\tau_{max}/\tau_{II}}\geq 1/2
\end{equation}%
If the above inequality is satisfied, then it is possible to build a
quantum memory which is able to store data with arbitrary success
probability over arbitrarily long times whereby the resource
scalings involved are of the form described in the earlier sections
of this article. The only implication of properly considering memory
errors in the derivation of the 2-trees resource scaling is that the
degree of the polynomial dependence on the tree branching parameters
will change. Properly considering memory errors effectively reduces
the success rate of the Type-II fusion gate by (at worse) a factor
of $P_{\tau_{II}} ^{\left( 2\tau_{max}/\tau_{II}\right)}$ as such
errors can be absorbed in the Type-II fusion gate as loss errors.
This in effect would increase the degree of the polynomial
dependence the 2-trees overhead has on the tree branching parameters
(see Section 3). On the other hand the proper consideration of the
memory errors during the building process of the quantum memory has
no effect on the derivation of $\tau_{max}$, the maximum time
individual photons need to be stored for in the process of building
and using the quantum memory proposed in this article.

Let us give an example with some sensible values of the various
parameters involved, to give an idea as to what the expectations are
for $P_{\tau_{II}}$. Suppose that the detector efficiency, $\eta
_{D}$, and the source efficiency, $\eta _{S}$, are both 95\%.
Further assume that we have $P_{\tau_{II}} ^{\left(
\tau_{max}/\tau_{II}\right)}=85\%$. This means that the loss rate of
the initial 3-qubit GHZ states (and all the subsequent trees
produced using Type-II gates) using the linear optics circuit
proposed by \cite{varnava2} would be approximately 30\%. Further
suppose that we desire to implement a loss tolerant quantum memory
gate which will have an effective success probability: $P_{mem}\geq
$ 99.99\%. This probability is the combined probability of
successfully joining an encoded 2-linear cluster to a single data
qubit, and being able to perform the two logical X measurements. To
achieve this, it would suffice to create trees that have a success
probability of 99.999\% for performing a single qubit measurement on
a tree encoded qubit \cite{varnava} and to create the hypertrees
involved with enough node photons such that the effective success
probability for joining two of them together would be 99.999\%.
(This is because $\left[ 99.999\%\right] ^{5}\geq 99.99\%)$.

The trees that can suppress a loss rate of 30\% to an effective
success probability of 99.999\% for performing the single qubit
measurement on tree encoded qubits have branching parameters
$\left\{ 11,23,22,4,1\right\} $ (data from \cite{varnava}). Each
Type-II gate will succeed with probability $\geq
P_{II}(0.85)^2\simeq$ 14.5\% with the values of $\eta _{D}$ , $\eta
_{S}$\ and $P_{\tau_{II}} ^{\left( \tau_{max}/\tau_{II}\right)}$
given above. Thus $k$, the number of node photons that have to be
present to boost the effective probability of joining hypertrees
together, $P_{CZ}$, to 99.999\% is\ $\sim 74$. The number of bonds
each hypertree forms with other hypertrees, $n,$ is equal to 2 since
we are only building linear cluster states for the needs of the
proposed memory (c.f. in a linear cluster state each qubit is at
most connected to 2 other qubits).

Substituting these values in the expression for $\tau_{\max }$ we
find that the number of $\tau _{II}$ time steps which are required
for the memory 2-qubit linear cluster are $\sim 25$. Therefore we
require that:
\begin{equation}
 P_{\tau_{II}} ^{25}=0.85
 \Rightarrow P_{\tau_{II}}=0.993
\end{equation}%

Therefore in this specific example we demonstrated that logical
qubits can be stored for a time of $25\tau _{II}$  with a success
probability of $\geq 99.99\%$ provided that individual photons can
be stored for a time $\tau _{II}$ with probability of 99.3\%
(assuming of course the values given for the detector and source
efficiencies as well).
\begin{figure}[ht]
\centering
\includegraphics[width=12cm]{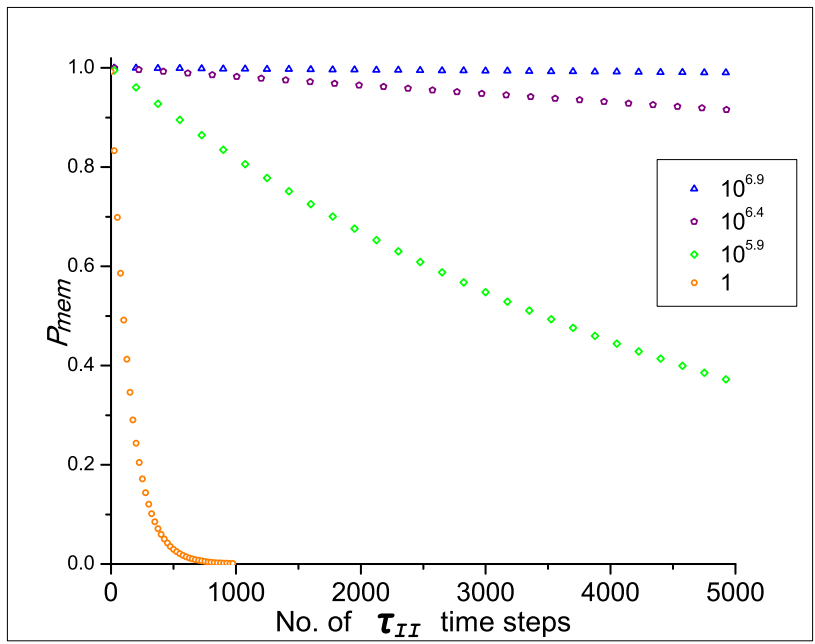}
\caption{\label{datafig} Graph showing how $P_{mem}$ varies with
storage time when the tree-encoded memory is implemented. The legend
gives the number of qubits present in the hypertrees for making up
the tree-encoded, memory cluster states for each curve.}
\end{figure}
Comparing with technology which is currently available we see that
the value of $P_{\tau_{II}}$ derived above is a bit demanding, some
2 orders of magnitude away from what is currently possible. For
example the cyclical quantum memory for photons proposed in
\cite{franson} has a cycle time of 13.3ns during which the
probability of successfully storing the the photon is 81\%. More
recently in \cite{zeilinger} it was shown that gate operation times
with active feed-forward take $\sim$ 150ns. Setting $\tau_{II}$ to
150ns shows that individual photon memory times should improve by at
least an order of magnitude in storage times and at least an order
of magnitude in the success probability rate in order to be able to
implement the proposed quantum memory.

In Fig.~\ref{datafig} we show how $P_{mem}$ can be affected by
simply varying the resources used in the tree encoded memory should
this value of $P_{\tau_{II}}$ be achieved. In each of the plots we
assume that the probability for storing an individual photon over
time $\tau_{II}$ is taken to be 99.3\% and observe how $P_{mem}$
varies when the number of qubits present in hypertrees is increased.
As we can see from Fig.~\ref{datafig} for the case when no encoding
is used, $P_{mem}$ drops to zero very rapidly in a time less than
1000$\tau_{II}$. However by increasing the number of qubits used in
hypertrees one can actively reduce the rate by which $P_{mem}$
decays. As long as Eqn.~(\ref{ineq}) is satisfied, then the decay
rate can in principle be reduced arbitrarily close to zero.

\section{Conclusion}

In this article we showed that it is possible to loss tolerantly
create a quantum memory based on a teleportation-type method which
itself is tolerant to photon loss. The method exploits the fact that
successive pairs of measurements of qubits in the X-basis in linear
cluster states have the effect of performing the identity gate. We
demonstrated that the success probability with which data qubits can
be stored with can approach unity exponentially fast by polynomially
increasing the resource overhead with respect to the success
probability. We also showed that the resources only need to scale
polynomially with respect to the time we wish to keep a qubit
stored.

In addition we showed that the maximum time required to store
photons in order to create an elementary unit of the the loss
tolerant memory - namely the 2-qubit linear cluster state - is
logarithmically dependent on the resources required. Strictly
speaking, this can indeed destroy the threshold result, however,
from a practical point of view, this is a mild limitation since it
only affects storage for extremely long times.

In the scheme for the quantum memory we are proposing, we introduced
special cluster state structures (we called them hypertrees) which
allow the probabilistic Type-II gates to be used to perform logical
CZ-gates amongst tree-encoded qubits in a near-deterministic
fashion. Since it is straightforward to convert parity measurements
to entangling gates (see e.g. \cite{divincenzobeenakker,pittman}),
this raises the possibility of using these gates to implement an
additional layer of encoding for tolerance to more general errors,
while retaining the much relaxed loss threshold that our protocol
provides.

\ack

This research was supported by  DTO-funded U.S. Army Research Office
Contract No. W911NF-05-0397, Merton College, Oxford and the
Engineering and Physical Sciences Research Council (EPSRC) and the
EPSRC's QIPIRC.

\end{document}